\documentclass[11pt]{article}   
\usepackage{amsmath}
\usepackage{amsfonts}
\usepackage{amssymb}
\usepackage{latexsym}
\usepackage{epsf}

\usepackage{graphicx}
\reversemarginpar \textwidth  167mm \textheight 220mm
\def\Journal#1#2#3#4{{#1} {\bf #2}, #3 (#4)}

\def\NPB{{\em Nucl.~Phys.} B}
\def\PLB{{\em Phys.~Lett.} B}

\def\PRL{\em Phys.~Rev.~Lett.}
\def\PRD{{\em Phys.~Rev.} D}
\def\PRC{{\em Phys.~Rev.} C}
\def\PRep{\em Phys.~Rep.}

\def\YAF{\em Yad.~Fiz.}
\def\SJNP{\em Sov.~J.~Nucl.~Phys.}

\def\Acta Pol{{\em Acta~Phys.~Pol.} B}

\begin{document}

\title{\bf Screening and Anti-Screening Effects \\
in $J/\psi$ Production on Nuclei}
\author{K.G.~Boreskov\footnote{boreskov@heron.itep.ru}~,
\ A.B.~Kaidalov\footnote{kaidalov@heron.itep.ru}\\
{\small Institute of Theoretical and Experimental Physics, Moscow,
Russia}}
\date{ }
\maketitle
\begin{abstract}
Nuclear effects in $J/\psi$ hadro- and electroproduction on nuclei are considered
in framework of reggeon approach. It is shown that screening regime which holds
for electroproduction at $x_F \gtrsim 0.7$ and for hadroproduction at
$x_F \gtrsim - (0.3\div 0.4)$ is changed with anti-screening regime for smaller $x_F$ values.
\end{abstract}

Heavy-quark production on nuclei provides an important information
on strong-interaction mechanisms and is intensively discussed (see e.g.
refs \cite{Vogt99,Vogt00,kopel} for review of experimental data,
discussion of some phenomenological models and more references).
In this note we will discuss a phenomenon of changing screening regime to
anti-screening one in $J/\psi$ hadro- and electroproduction when $x_{\psi}\equiv x_F$
decreases
in the framework of the BCKT model \cite{BCKT,Boreskov94} based on the reggeon approach.
For $J/\psi$ hadroproduction it happens in the region of negative $x_{\psi}$.
An investigation of this region can provide an important information on dynamics
of charmonium production and may discriminate different dynamical models
\cite{Vogt00,BCKT,Koudela03}.

Nuclear effects are usually discussed in terms of conventional
power-law parameterization
$F_{A}(x_{\psi})\propto F_{N}(x_{\psi})\cdot A^{\alpha (x_{\psi})}$,
where $F_{N}(x_{\psi})$ ( $F_{A}(x_{\psi})$ ) is inclusive $J/\psi $ production cross
section on a nucleon (on a nucleus). The function $\alpha (x_{\psi})$ characterizes
nuclear effects at different longitudinal momentum fraction $x_{\psi}$.

Experimental data on $J/\psi$ hadroproduction reveal a striking
contradiction with simplest theoretical expectations. Experimentally \cite{E772,E866,NA50}
function $\alpha (x_{\psi})$ decreases from $0.93\div 0.95$ at $x_{\psi}\simeq 0$
to values $\sim 0.75$ at $x_{\psi}\simeq 0.8$ thus indicating an increase of
the absorption as $x_{\psi}$ increases. Formation-time mechanism predicts
an opposite behavior.
Instead of expected scaling with $p_{J/\psi}^{lab}$ experiment shows
an approximate Feynman scaling with $x_\psi$ \cite{E866}.
Comparison of $J/\psi$ production data at different energies exhibits
an explicit breakdown of QCD factorization theorem for this process \cite{VBH}.

On the other hand when comparing $x$ dependence of $\alpha$ for $J/\psi$
hadroproduction with one for light-quark particle hadroproduction
(inclusive production of pions, nucleons, lambdas, etc.),
one observes the same trend -- a small absorption at $x\simeq 0$,
large absorption at $x\simeq 1$ and approximate Feynman scaling.
The difference is only
quantitative one -- $\alpha $ is about \hbox{$0.4 \div 0.5$} at $x\simeq 1$
for light quarks instead of $\sim 0.7$ for heavy quarks.
This behavior of $\alpha (x)$ for light hadrons allows a natural
explanation in Regge theory. Small absorption at $x\sim 0$ is due to
Abramovskii--Gribov--Kancheli (AGK) cancellation for inclusive spectra
\cite{AGK}, and increasing absorption at high $x$ is due to violation
of AGK rules because of momentum conservation requirement \cite{capella76}.

In ref.\cite{BCKT} the model for heavy quark and lepton pair production was
constructed taking into account the most essential aspects of reggeon approach.
In the spirit of parton picture \cite{Gribov73} it was suggested that a fast
projectile looks due to quantum-mechanical fluctuations like a cloud of
virtual particles consisting of light partons (quarks and gluons, or light-quark
hadrons, mainly pions) and (with some small probability) of heavy partons (say
$c\bar{c}$ pair).\footnote
{~In reggeon theory a mechanism of emitting these partons is supposed
to be a multiperipheral one. Creation of heavy  quarks in this model can be
considered as an example of "intrinsic charm" mechanism which was discussed by
Brodsky et al. \cite{intch,VBH}. In reggeon approach one can  consider
on equal footing any heavy component (e.g. ``an intrinsic lepton pair'' and so on).}
Different constituents of this fluctuation interact with nuclear matter
and determine a dependence upon atomic number $A$. There are both elastic and
inelastic interactions which from the viewpoint of reggeon diagrams are the
different discontinuities of the same reggeon diagram. As a result there
exist definite numerical relations between the discontinuities of different
types (AGK rules \cite{AGK}).

As it was stressed in ref.\cite{BCKT} one has to distinguish two types of reggeon
diagram cuttings.  For diagrams of the first class the registered particle is
contained {\em inside} a pomeron and appears in intermediate state
only due to cutting of this pomeron. In this case the AGK rules are always valid
and all contributions of many-pomeron diagrams to inclusive spectra cancel. It happens because each
additional pomeron can be both cut and uncut, and due to the opposite signs
of inelastic and elastic rescatterings they cancel. As a result, only one-pomeron
diagram contribution survives giving a linear A  dependence of spectra. This situation is
typical for low $x$ particles.

Another situation occurs if a particle to be registered is contained inside
``vertex'' of reggeon diagram (we use a term ``vertex'' for part of diagram common for
several pomerons attached to it). In this case AGK cancellation is not valid in general.
The striking manifestation of this mechanism comes from $A$ dependence of
inclusive spectra at large $x$ close to 1 (momentum conservation mechanism
\cite{capella76}).
The fastest particles belong to the ``vertex'' but not to the pomeron, and corresponding
discontinuities give $\alpha\approx 1/3$ for large $A$ compared to
$\alpha\approx 1$ for small $x$.

Similar effect exists if one observes the ``vertex'' particle of some particular
type, e.g. the $J/\psi$ meson. There are two sources of nuclear effects in this case
\cite{BCKT}. The first one is connected with rescatterings of light partons
in fluctuation on different target nucleons or, alternatively in the anti-laboratory
frame, with fusion of fluctuations originated from different nucleons
of fast nucleus. This mechanism plays a small role at present energies
$p_{lab}\lesssim 10^3~GeV$ \cite{BCKT} and its contribution is inessential for anti-screening
effects we are interested in.

Another contribution is connected to rescatterings of the charmed state itself.
One of the important ingredient of the model of ref.\cite{BCKT}
is an assumption that it is not the $J/\psi$ meson that propagates along nucleus but
some primary colorless system
containing both $c\bar{c}$ quarks (in color state) and light quarks (to screen
this color charge). This state denoted as $X$ can be $D\bar{D}$ mesons, or
$D^{\ast}\bar{D}^{\ast}$ state or something else, but the crucial point
is that this system is of large size and therefore strongly interacts with
nucleons (with cross section of order of $20\div 30 ~mb$). Only at last stage this
system $X$ converts into $J/\psi$ meson (or in $\psi'$) which is registered experimentally.

After inelastic rescattering the state $X$ can be lost for registration
(e.g. due to conversion to a state with small projection into $J/\psi$).
One can describe this effect quantitatively introducing a probability $\epsilon$
to find the state $X$ after its inelastic rescattering.
Then for $\epsilon=1$  one has exact AGK cancellation,
for $\epsilon=0$ the situation  is similar to momentum conservation mechanism
for light mesons at $x$ close to 1 (no AGK cancellation) with $\alpha\simeq 1/3$, and
for $0 < \epsilon < 1$ the situation is an intermediate one with incomplete AGK
cancellation. Moreover, the momentum of the state $X$ decreases after each
inelastic rescattering.This redistribution of the longitudinal momentum of
the state $X$ is essential for anti-screening effects discussed below.

Let us denote the momentum fraction distribution of system
$X$ immediately after its production as $F_{0}(x)$. The state $X$ has
some projection into state $J/\psi$ (as well as into $\psi'$, etc.)\footnote
{~For simplicity we suggest that only one state $X$ has a
noticeable projection into $J/\psi$. In order to describe properly the observed
difference of nuclear dependencies for $J/\psi$ and $\psi'$, one has to consider
at least two primary states, $X_1$ and $X_2$, with different projections to
$J/\psi$ and $\psi'$ and slightly different interactions with nuclear matter.
}.
The $x_{\psi}$ distribution can be obtained from $F_{0}(x)$ by convolution
with some projection function $G_{\psi}$:
\begin{equation}
F^{(\psi)}_{N}(x_{\psi}) = F_{1} \otimes G_{\psi}
\end{equation}
where the notation was used
\begin{equation}
(f\otimes g)(x)=\int_{x}^{1} \frac{dz}{z}f(z)\,g(x/z) ~,
\end{equation}
or, in the rapidity variables,
\begin{equation}
(f\otimes g)(\bar{y})=\int_{0}^{\bar{y}} d\xi f(\xi)\, g(\bar{y}-\xi) ~,
\end{equation}
where $\bar{y}=Y-y$ is a rapidity in the anti-lab frame.

The state $X$ travelling through nuclear matter suffers both elastic and
inelastic rescatterings. As in ref. \cite{BCKT}  the change of $x$ distribution
due to inelastic rescattering will be described as a convolution with some
function $\epsilon G(z)$, where $G(z)$ is normalized $(\int\limits^{1}_{0}
G(z)dz/z=1)$, and the parameter $\epsilon$ determines a probability to have
again the state $X$ after inelastic interaction (though with different momentum).
After $k$ inelastic rescatterings of the state $X$ we have $k$-fold convolution:
\begin{equation}\label{conv_k}
F^{(\psi)}_{k} =F_{1}\otimes {\underbrace{\epsilon G \otimes\cdots\otimes
\epsilon G}_{\text{$k$ times}}} \otimes G_{\psi}~~,
\end{equation}
It is essential that the operation of convolution is commutative and associative,
so we can first convolute $F_1$ with $G_{\psi}$ in the Eq.(\ref{conv_k}) and thus
get the $k$-fold convolution of the $J/\psi$ spectrum on the nucleon,
$F^{\psi}_{N}(x)$, with rescattering functions $G$:
\begin{equation}\label{F_k}
F_{k}^{\psi}= F^{\psi}_{N}\otimes {\underbrace{\epsilon G \otimes\cdots\otimes
\epsilon G}_{\text{$k$ times}}} ~.
\end{equation}

In order to obtain the $J/\psi$ inclusive spectrum on a nucleus, the functions $F^{(\psi)}_{k}$
should be weighted with cross sections $\sigma_{A}^{(k+1)}$ for $k$ inelastic rescattering
after $X$ production:
\begin{equation}\label{psi_k}
F^{\psi}_{A}=\sum^{\infty}_{k=0} F_{k}(x_{\psi})\, \sigma_{A}^{(k+1)} ~.
\end{equation}
The explicit form of the weights $\sigma_{A}^{(k+1)}$ depends upon energy region \cite{BCKT}.
However the change $\sigma_{A}^{(k+1)}$ is rather smooth and for simplicity we will use
the same low-energy expressions as in ref.\cite{BCKT}:
\begin{equation}\label{sigma_k}
\sigma_{A}^{(k+1)} = \sigma_{aN}^{(\psi)} \int d^2b \int_0^{T_A(b)} dv
\exp(-v \sigma_X)(v\sigma_X)^k/k! ~,
\end{equation}
where $\sigma_{aN}^{(\psi)}$ is the cross section of $J/\psi$ production on a nucleon,
$\sigma_{X}$ is the cross section of interaction of the state $X$ with a nucleon, and
$T_{A}(b)$ is two-dimensional nuclear density as a function of impact
parameter $b$ (nuclear profile).
These formulas correspond to rescatterings ordered in longitudinal direction
(low-energy regime) and after summation over $k$ give well known
optical-type formula for the production cross section.
Thus we come to Eq.(\ref{psi_k}) for $J/\psi$ nuclear inclusive cross section
where $F_{k}(x)$ and $\sigma_{A}^{(k+1)}$ are defined by Eqs.(\ref{F_k}) and (\ref{sigma_k}).

Let us discuss parameterizations of distribution functions entering the model.
The function $F^{\psi}_{aN}(x_\psi)$ can be determined from experimental data on
$J/\psi$ production on nucleon. Its form strongly depends on a type of a
projectile $a$. For hadron beam it can be parameterized as
\begin{equation}
F^{\psi}_{aN}(x)=C_a (1-x)^{\beta_a} ~,
\end{equation}
where $\beta_{\pi}\simeq 2$ for pion and $\beta_p \simeq 3$ for proton beam.
For $J/\psi$ production by photons (and electrons) the form of this function is strongly
different: produced  $\psi$ particles are concentrated near $x_{\psi}\simeq 1$,
and $x_{\psi}$ distribution can be  parameterized in a power-like way
\begin{equation}
F^{(\psi)}_{\gamma N}(x)=\beta_{\gamma} x^{\beta_{\gamma}}~,
\end{equation}
with $\beta_{\gamma}\geq 6$. We choose the value $\beta_{\gamma}=6$.

As to the function $G(x)$, it is natural to suggest its similarity to the
function $F^{(\psi)}_{\gamma N}(x)$. For simplicity we choose exactly the same form
\begin{equation}
G(x)=\beta x^{\beta} ~,
\end{equation}
with $\beta$ considered as a free parameter.

Terms of the series (\ref{psi_k}) in number of rescatterings $k$ decrease fastly
with $k$, and it is enough to take into account several first terms ($4\div 5$).
The corresponding convolutions at fixed impact parameter $b$ are estimated
analytically, and the $b$ integration was performed numerically.
We use for calculations the Woods-Saxon parameterization of nuclear density
with standard values of parameters taken from ref.\cite{Bohr}.

Before comparison of theoretical predictions with experimental data let us discuss
firstly quantitative character of absorptive corrections. As we discussed above
at small $x\approx 0$ we have AGK cancellation, and the extent of its violation
is proportional to $1-\epsilon$, a probability to lose the state $X$ after
inelastic rescattering. The particle can be lost for registration not only if it
disappears after interaction but also if it loses part of its momentum and is
shifted into different region in $x$. As a result at large $x$ values the AGK
rules are violated and absorption effects can be observed (momentum conservation
effect). However it is only
redistribution (not a loss) of particles for the whole $x$ region
(see Fig.1 for illustration of this fact).
For cross section integrated over $x$ screening (absorption) effects are connected only with
$1-\epsilon$ value but not with momentum conservation. For $J/\psi$ photoproduction
the absorption effects were usually analyzed only for integrated cross section
and this is a reason for rather small values of effective cross sections extracted
for $J/\psi$ absorption ($3\div 5~mb$). We want to demonstrate here \footnote
{~First this phenomenon was discussed in the talk \cite{Boreskov94} based on
unpublished results by K.~Boreskov and A.~Kaidalov.}
that as a function of $x$ nuclear effects can be rather big (with effective
cross section of order of $20\div 30~mb$) but integrally they cancelled
because of cancelling contributions from screening and anti-screening regions.

This is demonstrated at Fig.2 where we show description of
the NMC data \cite{ NMC}
on ratio $R(Sn/Be)$ of $J/\psi$ electroproduction cross sections as a function of
$x\equiv x_F$.
Since in this case the $J/\psi$ spectrum is concentrated at $x$ close to 1, the
anti-screening regime takes place already at $x \lesssim 0.8$. Note that
theoretical predictions at small $x$ value ($x < 0.6$) are less reliable due to
their sensitivity to small-$x$ parameterization of functions $F_{\gamma N}^{(\psi)}$
and $G(x)$.

For $J/\psi$ hadroproduction the anti-screening regime can be expected
(see Fig.1b) only for negative $x$. Fig.3 and Fig.6 show that for
$J/\psi$ and $\psi'$ hadroproduction the change of regime happens
at $x\approx -(0.3\div 0.5)$. This value depends on parameters of the model --
the less are momentum losses (the larger is $\beta$) and the less is $\epsilon$
the more negative $x$'s are necessary for the anti-screening onset (see Fig.3).
For example, for the value of $\beta=4$ the crossover happens already at $x\sim -0.2$.

It is important to note that it is not entirely adequate to analyze such
delicate effects as variation of dependence on atomic number with change of
$x$ or $p_T$ in terms of the standard parameterizations $A^\alpha$.
In general, $\alpha$ is not only a function of $x$ and $p_T$ but depends also
on atomic number $A$. To illustrate this fact we show at Fig.4 the ratios
of spectra $R(A/N)$ versus $A$ at fixed $x_{\psi}$
which by no means are described with a power dependence.
The same effect is demonstrated in Fig.3: there is a noticeable difference of
solid and dashed theoretical curves which correspond to different choices of
$A$ ranges.
Therefore the $\alpha$ value extracted from experiments depends significantly
on a chosen range of $A$ and on a way of analysis.
To exclude this uncertainty it is preferable to compare with theory ratios of cross
sections for different nuclei. Description of such data from E772 experiment
is shown in fig.5.

A difference in nuclear effects in $J/\psi$ and $\psi'$ hadroproduction
should be connected  in the model with presence of several primary states,
say $X$ and $X'$, which interact with nuclear matter in different ways and
have different branchings into $J/\psi$ and $\psi'$.
In Fig.6 we give as an example a comparison of data on $\psi'$ hadroproduction
from refs \cite{E772,E866} with model calculation with a single $X$ state at
the same parameter values as in fig.2.
The two-channel generalization is evident.

Let us mention that in some theoretical models \cite{Vogt00,Koudela03} used for
description of $J/\psi$, $\psi'$ hadroproduction in the negative $x$ region
the function $\alpha (x)$ decreases (screening effects increase) as $|x|$
increases in this region.

Thus, an experimental investigation of nuclear effects for charmonium production
in the whole region of $x$ can provide a valuable information for testing different
dynamical models. In the reggeon approach one can distinguish three different
regions in rapidity or in Feynman variable for $J/\psi$ hadroproduction --
in central region one has strong cancellation of screening diagrams and $\alpha$
is close to one (small absorption cross section); at $x > 0.2$ due to violation
of cancellation rule the absorption effects increase and effective cross section
is about $20\div 30~mb$; at negative $x$ less than $-(0.3\div 0.4)$ the anti-screening
regime should be observed.
Data on $A$ dependence of charmonia production in the negative $x$ region will be
available soon from HERA-B experiment \cite{HERA-B}.
\\[3mm]
{\bf Acknowledgements.}
{\small We are grateful to A.~Capella and M.~Danilov for stimulating discussions.
This work has been partially supported by RFBR grants 01-02-17383 and 00-15-96786,
INTAS grant 00-00366 and DFG grant 436 RUS 113/721/0-1.}

\small{

}
\newpage
\begin{figure}[h]\label{resc}
\begin{center}
\epsfxsize=10cm
\epsfysize=15cm
\epsfbox{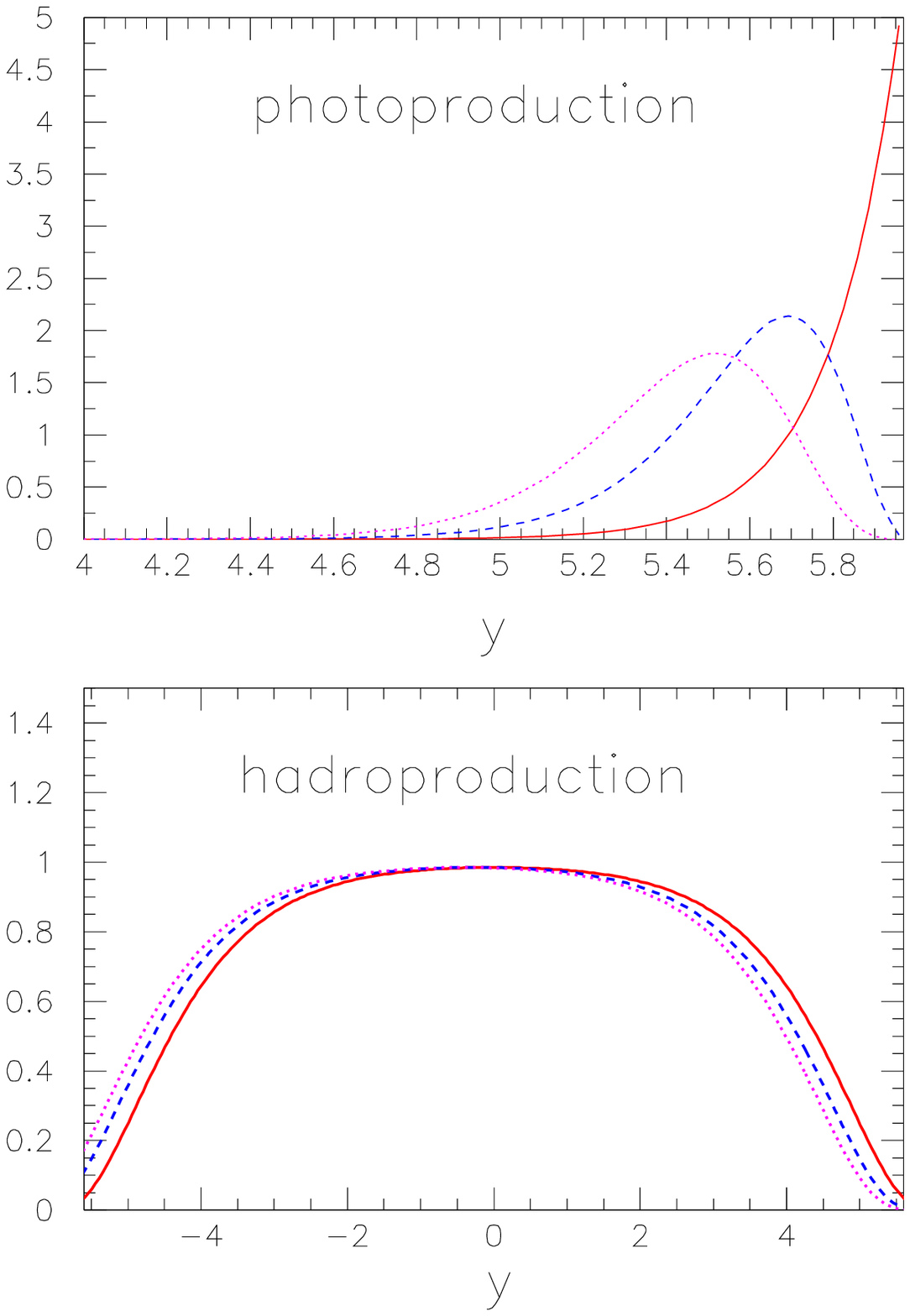}
\end{center}
\caption{Change of the original spectrum (solid curves) after 3 (dashed curves)
and 5 (dotted curves) rescatterings for: (a) photoproduction
and (b) hadroproduction of $J/\psi$.}
\end{figure}
\begin{figure}[h]\label{gamma}
\begin{center}
\epsfxsize=15cm
\epsfbox{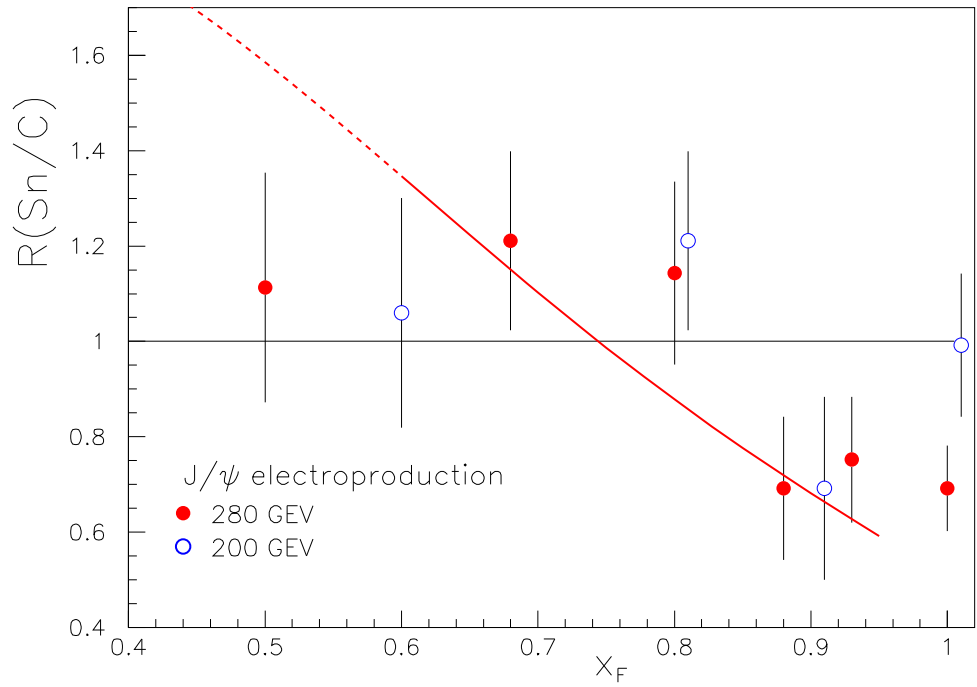}
\end{center}
\caption{Description of data \cite{NMC} on ratio $R(Sn/C)$
for $J/\psi$ electroproduction.
The curve corresponds to model calculations at parameter values
$\sigma=30~mb$, $\epsilon=0.85$, $\beta = 12$.}
\end{figure}
\begin{figure}[h]\label{psi}
\begin{center}
\epsfxsize=15cm
\epsfbox{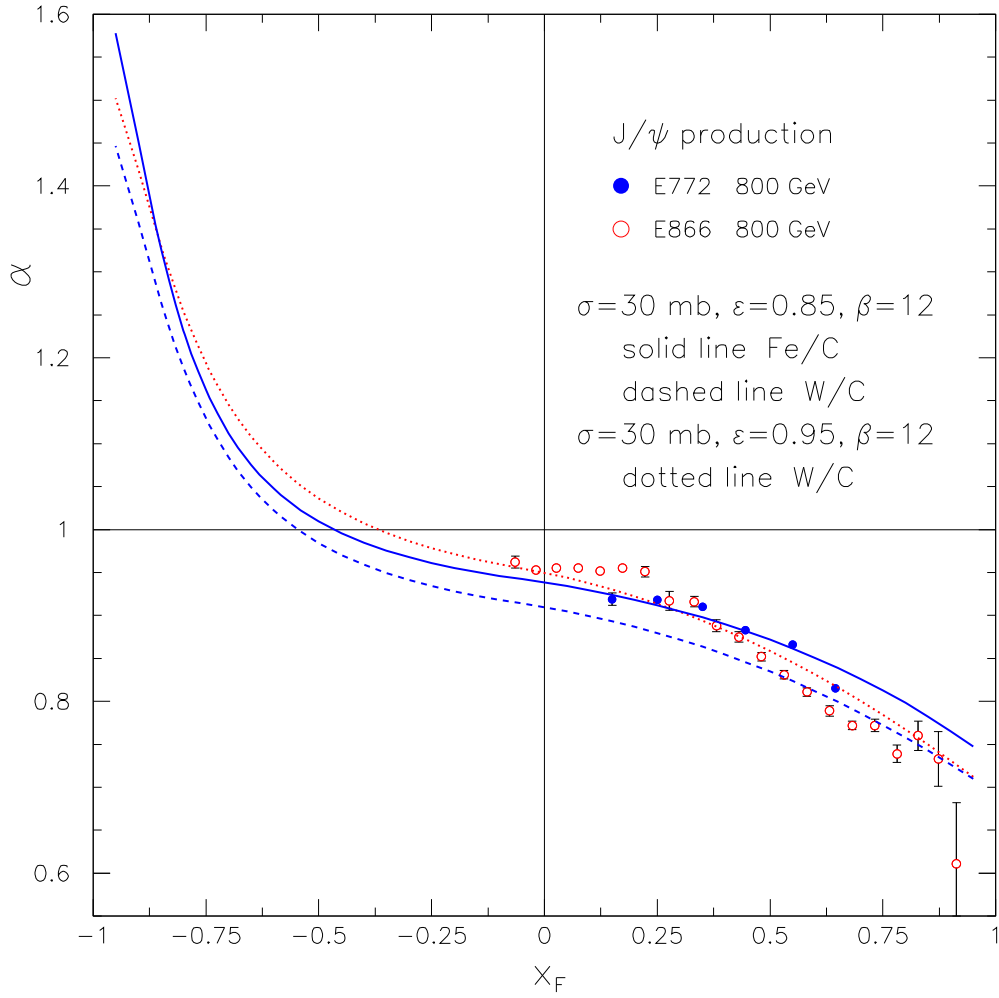}
\end{center}
\caption{Description of data \cite{E772,E866} on nuclear dependence
of $J/\psi$ hadroproduction.
Curves correspond to model calculations for different ways of $\alpha$ extraction:
solid curve -- from the ratio $R(Fe/C)$ and dashed curve -- from $R(W/C)$.
The dotted line extracted  from $R(W/C)$ corresponds to $\epsilon=0.95$.}
\end{figure}
\begin{figure}[h]\label{psir}
\begin{center}
\epsfxsize=15cm
\epsfbox{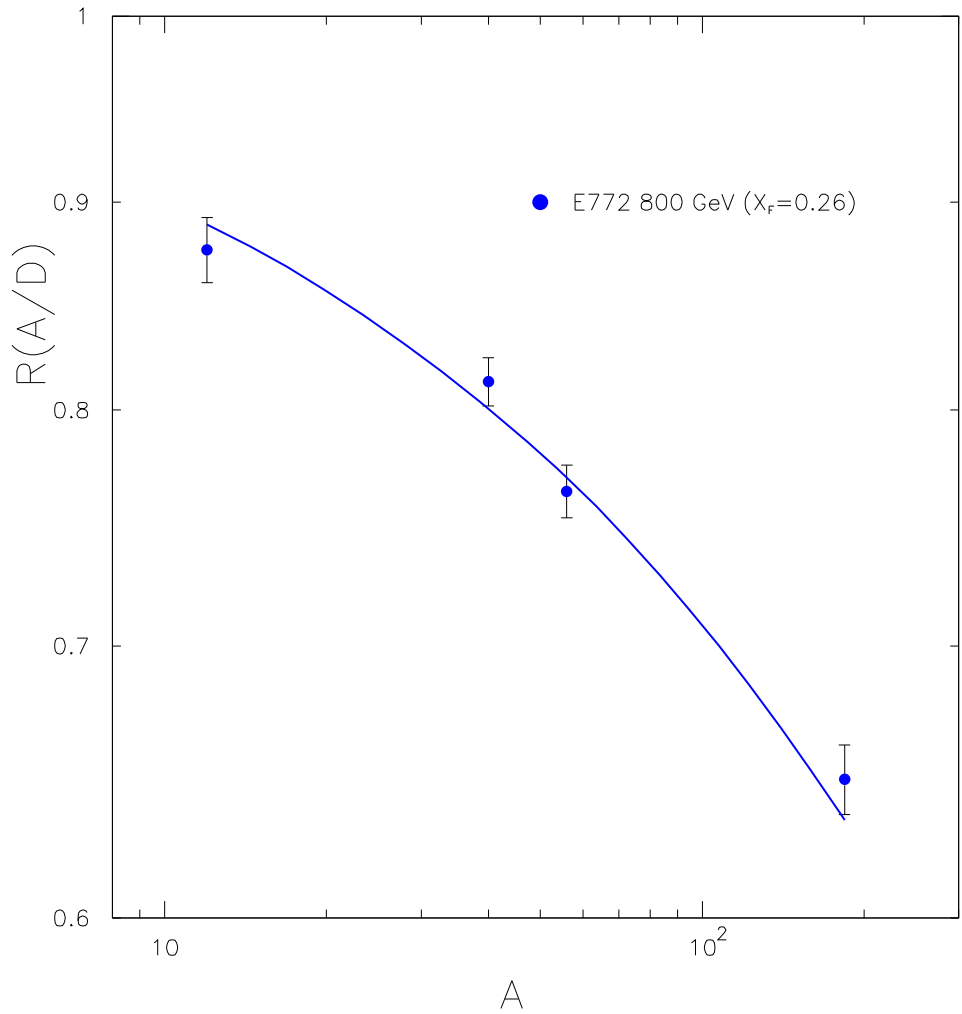}
\end{center}
\caption{Ratios $R(A/D)$ vs $A$ at fixed $x_F=0.26$.}
\end{figure}
\begin{figure}[h]\label{psirx}
\begin{center}
\epsfxsize=15cm
\epsfbox{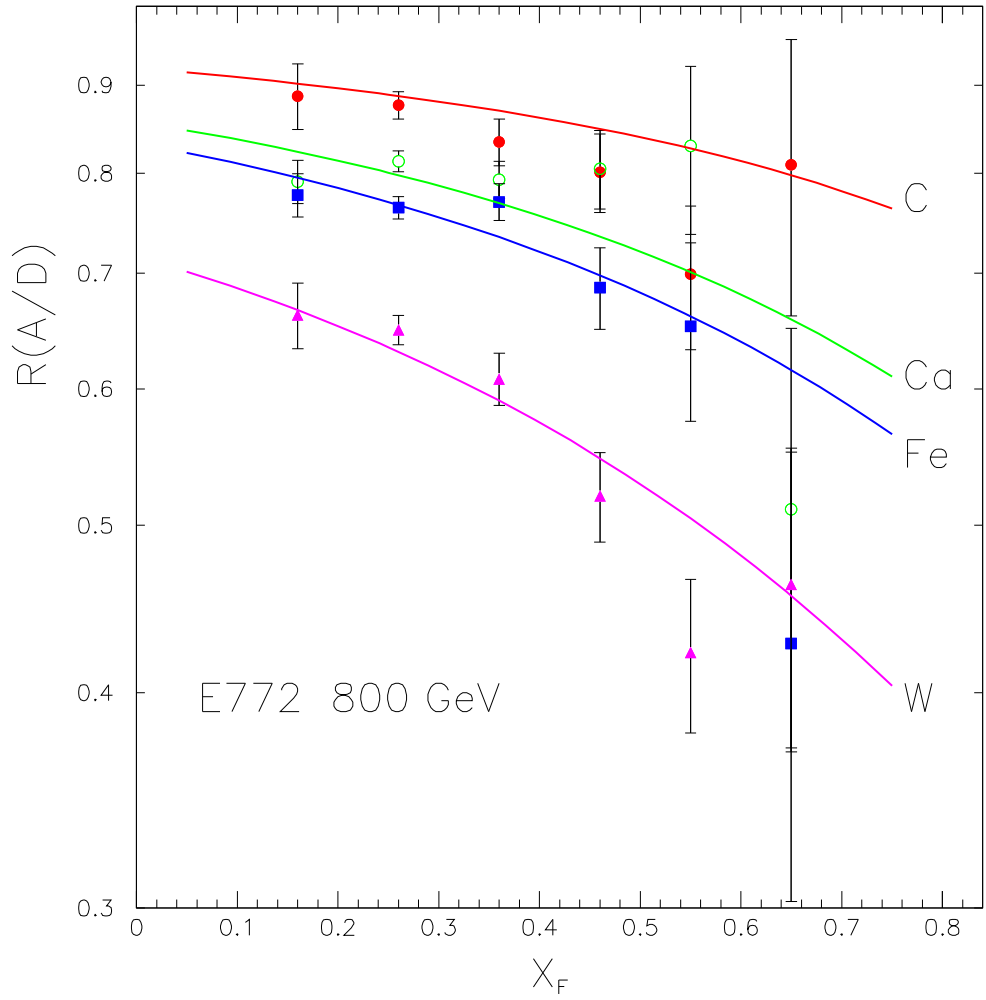}
\end{center}
\caption{Ratios $R(A/D)$ vs $x_F$ for different atomic numbers $A$. }
\end{figure}
\begin{figure}[h]\label{psip}
\begin{center}
\epsfxsize=15cm
\epsfbox{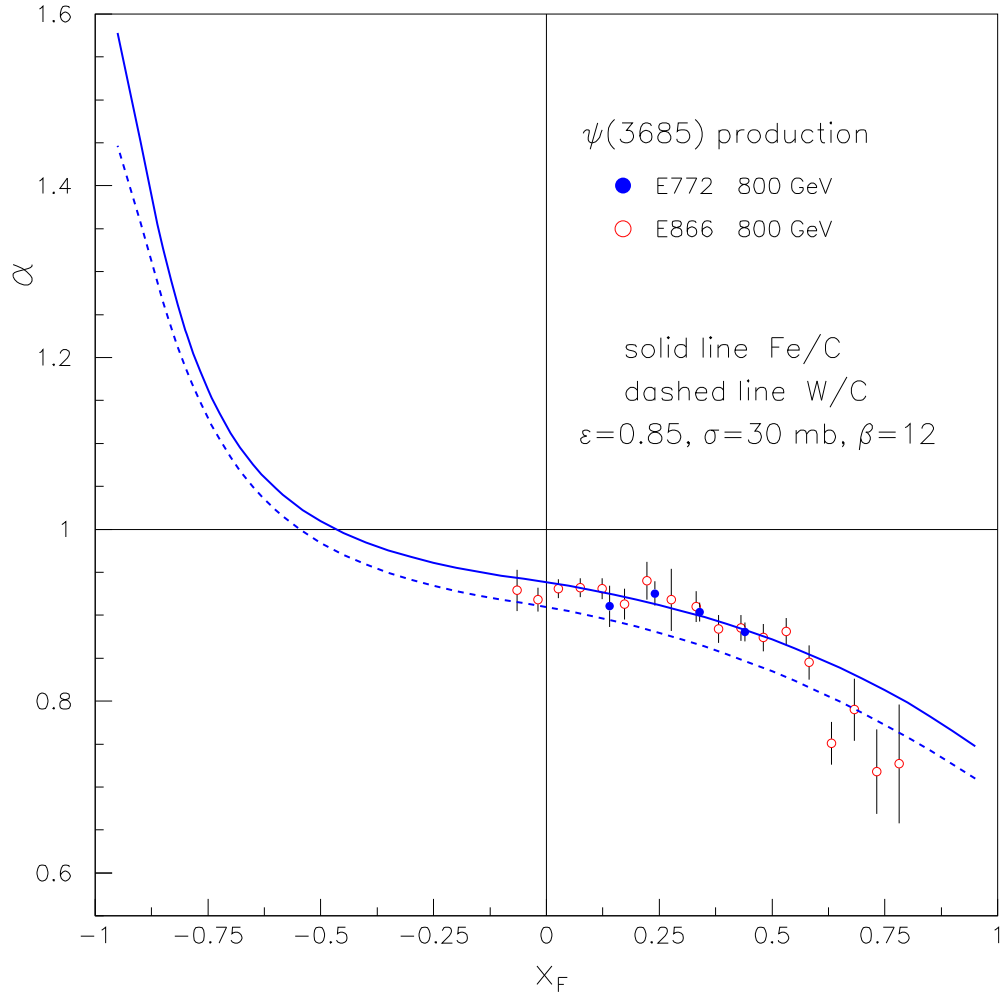}
\end{center}
\caption{Description of data \cite{E772,E866} on nuclear dependence
of $\psi'$ hadroproduction.
The curve corresponds to model calculations at parameter values
$\sigma=30~mb$, $\epsilon=0.85$, $\beta = 12$.}
\end{figure}
%
%

\end{document}